# Retrieval of multimedia stimuli with semantic and emotional cues: Suggestions from a controlled study

Marko Horvat[1], Davor Kukolja[2] and Dragutin Ivanec[3]

[1]Polytechnic of Zagreb, Department of Computer Science and Information Technology
[2]University of Zagreb, Faculty of Electrical Engineering and Computing, Department of Electric Machines, Drives and Automation
[3]University of Zagreb, Faculty of Humanities and Social Sciences, Department of Psychology
E-mail: marko.horvat@tvz.hr

**Abstract** - The ability to efficiently search pictures with annotated semantics and emotion is an important problem for Human-Computer Interaction with considerable interdisciplinary significance. Accuracy and speed of the multimedia retrieval process depends on the chosen metadata annotation model. The quality of such multifaceted retrieval is opposed to the potential complexity of data setup procedures and development of multimedia annotations. Additionally, a recent study has shown that databases of emotionally annotated multimedia are still being predominately searched manually which highlights the need to study this retrieval modality. To this regard we present a study with $N = 75$ participants aimed to evaluate the influence of keywords and dimensional emotions in manual retrieval of pictures. The study showed that if the multimedia database is comparatively small emotional annotations are sufficient to achieve a fast retrieval despite comparatively lesser overall accuracy. In a larger dataset semantic annotations became necessary for efficient retrieval although they contributed to a slower beginning of the search process. The experiment was performed in a controlled environment with a team of psychology experts. The results were statistically consistent with validates measures of the participants' perceptual speed.

## I. INTRODUCTION

Manual searching of multimedia datasets such as picture repositories and video databases is still predominately based on keyword-based information retrieval. However, range and expressivity of manual data description process are limited. When adding a picture to an Internet repository or social network users are asked to describe the picture, who or what is portrayed, when and where the photo was taken, which emotion does the picture stimulate etc. Frequently users provide contextual information which is insufficient for searching. The retrieved datasets are large and inaccurate [1]. Various semantic categories, available multimedia content and metadata are used to boost retrieval accuracy and improve ranking of results important to the posed query [2] [3] [4] [5]. These procedures are the subject of ongoing research but each of them has restrictions and in practice cannot completely substitute missing expert information.

The missing data problem is especially exacerbated in affective multimedia databases because they implement only sparse annotation with unrestricted keywords [6]. These databases are especially important tool for study of emotion and attention in the areas of psychology, neurology and cognitive sciences [7]. Retrieval of loosely annotated data, with only one expert-defined tag per picture, is even more open to errors than with, for example, social media with community generated tags, hashtags or folksonomies. Although tags are noisy they can still provide useful information if combined with intelligent algorithms for statistical reasoning [8]. As has been shown in a recent survey on usage modalities of emotionally-annotated picture databases [9], the ultimate consequence of such poor semantic annotation scheme is that it is more productive to browse affective multimedia databases visually, although they may contain hundreds or thousands of images, than to search them with keywords. Additionally, this survey showed that 70% of experts need more than 1h to retrieve optimal pictures for a single emotion elicitation sequence and 20% require more than 12h which is extremely impractical. Therefore, it is necessary to establish how much context information is enough to enable efficient keyword and emotion search paradigms in affective multimedia databases. How formal and expressive must the data representation be? Is it necessary to better describe multimedia semantics or emotion, or both? These and other similar questions may be theoretically examined but ultimately only a controlled experiment can provide a definite answer to them. Annotation quality and retrieval efficiency are in a mutual tradeoff. More thorough annotation comes at a cost of better data preparation, but will result in a more accurate search. The experiment described in this paper was motivated to explore optimal methods for annotation of semantics and emotion in affective multimedia databases.

The presented research had two goals; considering the results of the survey it was necessary to determine how fast and accurately experts can find a specific picture stimulus in ideal circumstances. The second goal was to establish minimal emotion and semantic annotation requirements which still enable an efficient searching of affective multimedia databases.

The remainder of this paper is organized as follows; Section 2 provides detailed background information and describes the experimental setup. Section 3 specifies and visualizes all obtained results, and the following Section 4 discusses the results from the perspective of the experiment's goal. Finally, Section 5 concludes the paper and outlines future work into this subject.

## II. METHOD

The controlled experiment was performed at the Department of Psychology of University of Zagreb, Faculty of Humanities and Social Sciences. A total of $N$=75 college students (10 males, 65 women) with an average age 19.49 years (*std* = 0.87) participated in the experiment.

All participants performed two tasks; Task 1: find five pictures with particular emotion, and; Task 2: find five pictures with specific emotion and semantics. The search criteria can be considered similar to emotion elicitation sequences typically used in real settings. All participants first performed Task 1 and then Task 2. The results help to estimate the average time to retrieve a stimulus on the basis of the set of semantic and affective criteria, and the number of errors that can occur at the same time.

In both tasks, participants had to, as quick as possible perform visual search tasks in the valence-arousal emotion plane to find the desired stimuli [10] [11]. However, to achieve an optimal search strategy in the second task it was necessary to first input search keywords to reduce the stimuli subset and leave only those pictures that match the Task 2.

A preselected subset of 250 picture stimuli from the International Affective Picture System (IAPS) [12] database was used for both tasks. In total 50 pictures (20% of the subset) matched the search criteria. In Task 1 the participants had to find IAPS pictures with values of emotion dimensions within a given region of valence-arousal plane (*valence* > 5.0 and *arousal* > 5.0), while in Task 2 the pictures had *arousal* > 5.0 and preset semantics (i.e. children, families, specific animals etc.). For Task 2 the participants were given a list of keywords they had to match with the presented stimuli. All tasks were designed with positive, i.e. non-null, results. The tasks' texts and the list of picture stimuli used in the experiment are available by contacting the first author.

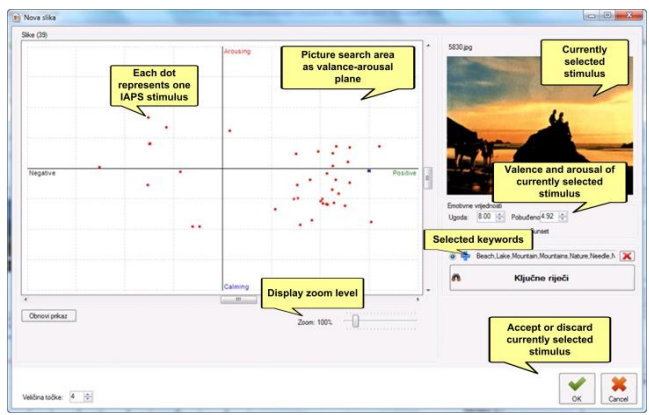

Fig. 1. The main screen of the "Intelligent stimuli generator" tool displaying IAPS stimuli subset in Task 2 defined with selected keywords.

The participants used a previously developed software tool "Intelligent stimuli generator" to search for stimuli with emotion and semantic parameters [7]. The tool was specially adapted for the experiment: user interface was greatly simplified for implementation of the experimental protocol. All search paradigms except the manual search and other unnecessary features were removed. However, a comprehensive logging module was added which can enumerate, record and timestamp different user actions, search attempts and categorize them as successful or not. The logs are generated automatically and can be later analyzed with dedicated statistical tools like SPSS or Matlab.

A week before the experiment the college students were given comprehensive written instructions about the "Intelligent stimuli generator" tool so they could familiarize themselves with the user interface and the types of tasks they will have to carry out. The experiment itself took place in the laboratory of the Department of Psychology of University of Zagreb, Faculty of Humanities and Social Sciences. The laboratory contains a number of separate sound-proof cabins with a bench, two seats, PC computer, 17" 4:3 LCD monitor and sound speakers. At a time each cabin seated one lab assistant with psychology training and one experiment participant.

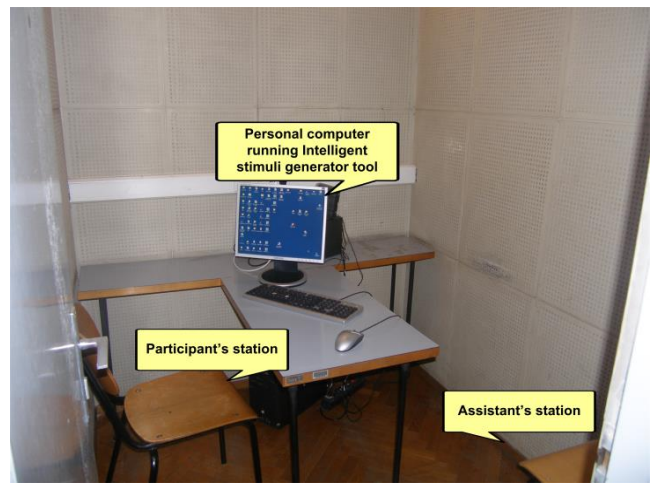

Fig. 2. A soundproof cabin used in the study with two stations for the participant and assistant, and a personal computer for picture retrieval with semantic and emotional cues.

Before the picture search experiment all participants took two standard psychological tests of their perceptual speed [13]. Each test lasted 5 minutes and they were also conducted in controlled conditions with minimum outside sensory interference. The aim of these additional tests was to investigate the relationship of participants' visual search performance and achievement on perceptual speed tests, as a measure of stable individual traits connected to the perceptual ability [14].

## III. RESULTS

Complete results of $N_{RTOT}$=750 pictures retrievals are shown in Fig. 3. Each point in diagrams represents an aggregated mean of $N_R$=75 records for each of the five searched pictures (numbered 1−5 on x-axis) in Task 1 and Task 2, respectively, with indicated standard deviation. Picture retrieval duration in seconds is provided on y-axis.

The results show that in Task 1 on average 10.74 s (*std* = 8.66) was necessary to find a picture using only valence and arousal emotion constraints. More time was needed to locate files by searching semantics. In Task 2 average 47.48 s (*std* = 15.18) was needed to retrieve a picture with specific keywords and emotion which is more than 4 times longer than the retrieval without semantics.

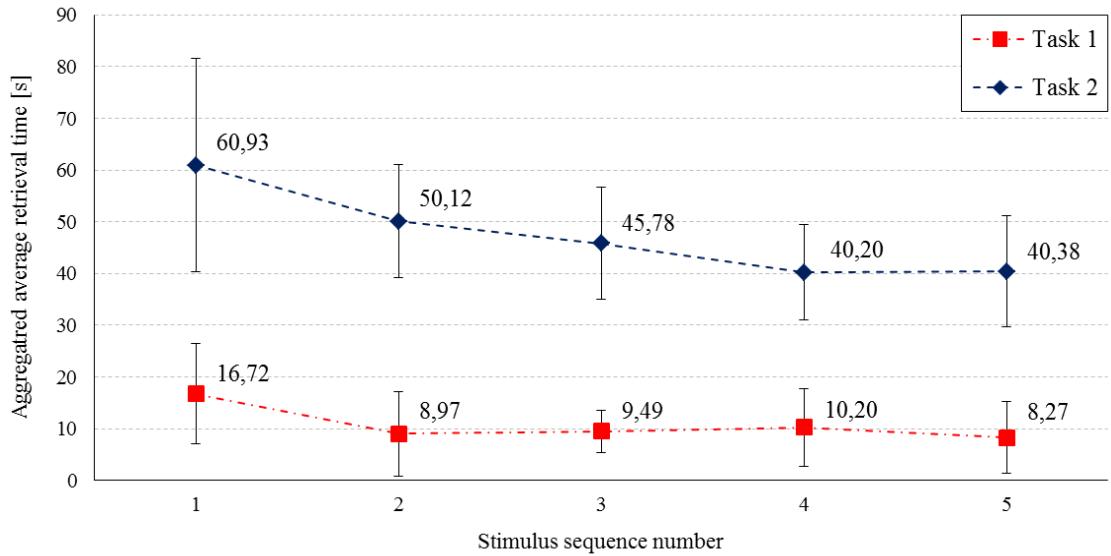

Fig. 3. Aggregated average performance of *N*=75 participants in Task 1 and Task 2. Individual searched pictures (labeled 1−5) are represented on x-axis, and the recorded search time is y-axis.

In Task 2 the search time monotonically decreases with each found picture and the difference between the first and the last search is over 20 s − from 60.93 s (*std* = 20.67) down to 40.38 s (*std* = 10.73). In Task 1 the average time spent to found one picture was marginally better with 16.72 s (*std* = 9.66) for the first and 8.27 s (*std* = 6.91) for the fifth and last picture.

In Table 1 are the correlations between the two perceptual speed, and Task 1 and Task 2. Both tests are significantly statistically related only with Task 2 but not with Task 1.

TABLE I
Correlation measures (*r*) among achievements on perceptual speed tests (Test 1, Test 2) and efficiency in picture search tasks (Task 1, Task 2). Each measure was calculated on $N_R$=75 sample (*$p<0.05$; **$p<0.01$).

|        | Test 1 | Test 2 | Task 1 | Task 2  |
|--------|--------|--------|--------|---------|
| Test 1 | 1      | 0.57** | 0.08   | -0.27*  |
| Test 2 |        | 1      | 0.11   | -0.33** |
| Task 1 |        |        | 1      | -0.07   |
| Task 2 |        |        |        | 1       |

The aggregated average number of incorrectly found images (false positives) in Task 1 was 2.75 (*std* = 2.75), while only 1.26 (*std* = 0.53) in Task 2 is an indication of greater search accuracy which was made possible by introduction of semantic constraints. Also, statically very large dispersal of false positives was present in Task 1 suggesting that some participants' search strategy was based on volume, i.e. fast processing of many pictures, rather than on accuracy and precision.

IV. DISCUSSION

The study showed that in ideal conditions, i.e. if the stimuli database is sufficiently small, it is possible to relatively quickly find a specific image just based on affective tagging. But at the same time this approach results in relatively large error rate and participants participant retrieved more false pictures before finding the correct one.

Stimuli retrieval with keywords was initially slower, which is especially apparent by comparing search times for first pictures in both tasks, but once the required keywords were found the displayed set of images contained enough true positives so the required picture could be retrieved relatively quickly and at the same time the participants consistently made less mistakes. In both tasks the average retrieval time asymptotically falls with the increasing sequence number of the image, indicating that respondents needed a period of getting used to work with the "Intelligent stimuli generator" tool, but they could quickly achieve their relative maximum search efficiency.

However, it should be pointed out that in this study the original IAPS database was substantially reduced and the participants searched a subset with only 250 of the original 1182 IAPS stimuli (21.15 %), and in this subset 96 pictures (38.4 %) matched the overall search criteria. It reasonable to assume that the search times and the number of errors being made would be grater if realistic (i.e. larger and more complex) emotionally annotated multimedia databases were used or if the participants could not have the benefit of having a pre-defined set of semantic tags at their disposal.

Perceptual speed is defined as the speed of accurate perception of details in observed structures. It is measured with the speed of accurate recognition of similarities and differences between the observed structures [14].

Correlation among scores on perceptual speed tests and Task 2 (stimuli retrieval with keywords) were obtained (Table 1). Correlations (i.e. Pearson coefficients) have negative values which can indicate that those participants who were better in perceptual tests required less time to find keywords given to them in Task 2. It also demonstrates that stable interindividual traits may, at least to some extent, explain efficiency of keyword-based database retrieval method. Additionally, it could also mean that an intelligence measure could explain part of retrieval efficiency, because the perceptual speed has low to moderate correlation with intelligence which points out to their common probable cause [15]. The obtained correlations are not strong, but stable − both tests have similar correlation to time of retrieval. Indirectly, correlation values indicate that when task was more complex some cognitive individual differences played a role in general retrieval efficiency. This was expected and means that participants were motivated in both tasks.

## V. CONCLUSION

Multimedia databases may be searched and retrieved by any combination of available text and visual features. With current knowledge models it is possible to describe emotion and high-level semantics in XML-derived formats thus enabling reasoning about emotional states, anxieties, phobias, stress or mental fatigue that a multimedia file may provoke or be related to [16]. During construction of affective multimedia retrieval information systems the optimal solution is to have the simplest possible requirements on the complexity of data model but at the same time to achieve high search efficiency and retrieval accuracy.

We showed that if emotionally annotated multimedia databases have a few hundred images it is not necessary to annotate their semantics. If databases are relatively small they can be efficiently searched just by using information about multimedia affective content. However, semantic-based search is practical in large multimedia sets. In a two-step retrieval process the search space must be first reduced with semantic constraints to a manageable size which enables multimedia to be searched efficiently with emotion or other metadata tags.

We hope that the presented study could be used in design of future affective multimedia retrieval systems as well as affective multimedia databases. The study's results may help researchers to find the optimal choice between the data model complexity and the required affective multimedia search efficiency.

## REFERENCES

[1] W. B. Croft, D. Metzler, D., and T. Strohman, "Search engines: Information retrieval in practice," Reading: Addison-Wesley, pp. 351−358, 2010.

[2] N. Rasiwasia, J. Costa Pereira, E. Coviello, G. Doyle, G. R. Lanckriet, R. Levy, and N. Vasconcelos, "A new approach to cross-modal multimedia retrieval," Proc. of the international conference on Multimedia, ACM, pp. 251−260, October 2010.

[3] Y. Yang, F. Nie, D. Xu, J. Luo, Y. Zhuang, and Y. Pan, "A multimedia retrieval framework based on semi-supervised ranking and relevance feedback," IEEE Trans. on Pattern Analysis and Machine Intelligence, vol. 34(4), pp. 723−742, 2012.

[4] S. Clinchant, J. Ah-Pine, and G. Csurka, "Semantic combination of textual and visual information in multimedia retrieval," Proceedings of the 1st ACM International Conference on Multimedia Retrieval, pp. 44:1–44:8, April 2011.

[5] D. Carmel, N. Zwerdling, I. Guy, S. Ofek-Koifman, N. Har'El, I. Ronen, ... and S. Chernov, "Personalized social search based on the user's social network," Proc. of the 18th ACM conference on Information and knowledge management, ACM, pp. 1227−1236, November 2009.

[6] M. Horvat, S. Popović, N. Bogunović, and K. Ćosić, "Tagging multimedia stimuli with ontologies," Proceedings of the 32nd International Convention MIPRO 2009, Croatia, pp. 203–208, May 2009.

[7] M. Horvat, N. Bogunović, and K. Ćosić, "STIMONT: a core ontology for multimedia stimuli description," Multimedia Tools and Applications, vol. 73(3), pp. 1103−1127, 2014.

[8] F. Abel, Q. Gao, G. J. Houben, and K. Tao, "Semantic enrichment of twitter posts for user profile construction on the social web," The Semanic Web: Research and Applications, Springer Berlin Heidelberg, pp. 375−389, 2011.

[9] M. Horvat, S. Popović, and K. Ćosić, "Multimedia stimuli databases usage patterns: a survey report," Proceedings of the 36nd International Convention MIPRO 2013, Croatia, pp. 993−997, May 2013.

[10] A. Mehrabian, "Pleasure-arousal-dominance: A general framework for describing and measuring individual differences in Temperament," Current Psychology, vol. 14(4), pp. 261−292, 1996.

[11] J. A. Russell, "A circumplex model of affect," Journal of personality and social psychology, vol. 39(6), pp. 1161−1178, 1980.

[12] P. J. Lang, M. M. Bradley, and B. N. Cuthbert, "International affective picture system (IAPS): Affective ratings of pictures and instruction manual," Technical Report A−8, University of Florida, Gainesville, FL, 2008.

[13] D. Ivanec, "New Perceptual speed tests. Unpublished tests developed under project: Development, standardization and psychometric validation of cognitive ability tests," Faculty of Humanities and Social Sciences, Zagreb, 2011.

[14] K. Rayner, "Eye movements and attention in reading, scene perception, and visual search," The quarterly journal of experimental psychology, vol. 62(8), pp. 1457−1506, 2009.

[15] L. D. Sheppard and P. A. Vernon, "Intelligence and speed of information-processing: A review of 50 years of research," Personality and Individual Differences, vol. 44(3), pp. 535−555, 2008.

[16] M. Schröder, P. Baggia, F. Burkhardt, C. Pelachaud, C. Peter, and E. Zovato, "EmotionML–an upcoming standard for representing emotions and related states," Affective Computing and Intelligent Interaction, Springer Berlin Heidelberg, pp. 316−325, 2011.